\documentclass{ws-ijmpb}

\begin{document}

\markboth{Authors' Names}
{Instructions for Typing Manuscripts (Paper's Title)}

%
\catchline{}{}{}{}{}
%

\title{Quantum discord under system-environment coupling: the two-qubit case}

\author{Jin-Shi Xu}

\address{Key Laboratory of Quantum Information, University of Science and Technology of China, CAS, Address\\
Hefei, 230026,
People's Republic of China}

\author{Chuan-Feng Li}

\address{Key Laboratory of Quantum Information, University of Science and Technology of China, CAS, Address\\
Hefei, 230026,
People's Republic of China\\
cfli@ustc.edu.cn}

\maketitle

\begin{history}
\received{Day Month Year}
\revised{Day Month Year}
\end{history}

\begin{abstract}
Open quantum systems have attracted great attention, since inevitable coupling between quantum systems and their environment greatly affects the features of interest of these systems. Quantum discord, is a measure of the total nonclassical correlation in a quantum system that includes, but is not exclusive to, the distinct property of quantum entanglement. Quantum discord can exist in separated quantum states and plays an important role in many fundamental physics problems and practical quantum information tasks. There have been numerous investigations on quantum discord and its counterpart classical correlation. This short review focuses on highlighting the system-environment dynamics of two-qubit quantum discord and the influence of initial system-environment correlations on the dynamics of open quantum systems. The external control effect on the dynamics of open quantum systems are involved. Several related experimental works are discussed.
\end{abstract}


\section{Introduction}
Interest in open quantum systems arises naturally from the fact that any quantum system is always unavoidably coupled to its surrounding environment\cite{Breuer2002}. This makes investigation of the preparation, processing, and application of features in open quantum systems both fundamentally and practically important. One of the most remarkable properties of quantum systems is the existence of correlations that do not have corresponding classical counterparts. Entanglement, a special kind of nonclassical correlation, is one such that has been widely studied\cite{Horodecki2009,Amico2008} and found to be a useful resource in quantum communications and quantum computation\cite{Bennett2000}. Entanglement is fragile, and significant effort has gone into investigating its behavior in open quantum systems\cite{Mintert2005}. One unusual dynamic entanglement behavior, entanglement sudden death, occurs when entanglement completely disappears at a finite point in the system's evolution\cite{Yu2009}. Recent studies have discovered quantum correlations other than entanglement, with some of these other nonclassical correlations even existing in separable quantum states. Quantum correlations without entanglement could potentially play an important role in implementing quantum information tasks that benefit from quantum advantages such as information locking\cite{DiVincenzo2004}, DQC1 (deterministic quantum computation with one quantum bit) algorithm applications\cite{Datta2008,Lanyon2008}, quantum state discrimination\cite{Roa2011,Li2012}, quantum metrology\cite{Modi2011}, and remote quantum state operations\cite{Dakic2012}. Quantum discord\cite{Ollivier2001} has also attracted significant attention as a way to quantify these total quantum correlations encoded in a quantum system\cite{Celeri2011,Modi2012}.

This short review focuses on the dynamics of two-qubit quantum discord under system-environment coupling, dividing the work into five sections. We first introduce the original definitions of quantum discord and classical correlation in Part 2, as well as introducing other popular related nonclassical measures. In Part 3, we review recent theoretical developments of two-qubit quantum discord under system-environment coupling, discussing the system's dynamic properties and external control effects. In Part 4, several related experiments are reviewed in detail, followed by a conclusion in Part 5.

\section{Quantum discord and classical correlation}

In classical information theory, the correlations between two different random variables $A$ and $B$, with probability distributions $p_{i}$ and $q_{i}$ corresponding to the obtained values of ${a_{i}}$ and $b_{i}$, can be characterized by mutual information\cite{Cover1991}:
\begin{equation}
I(A:B)=H(A)+H(B)-H(A,B), \label{mutual}
\end{equation}
In Eq. (\ref{mutual}) $H(A)$ ($H(B)$) is the Shannon entropy representing the uncertainty of the outcomes of $A$ ($B$), such that $H(A)=-\Sigma_{i}p_{i}\log_{2}p_{i}$ ($H(B)=-\Sigma_{i}q_{i}\log_{2}q_{i}$). Note that $H(A,B)=-\Sigma_{i,j}p_{ij}\log_{2}p_{ij}$ is the joint entropy where $p_{ij}$ represents the joint probability of the outcomes of $A$ and $B$ being $a_{i}$ and $b_{j}$, respectively, and $p_{i}=\Sigma_{j}p_{ij}$ ($q_{j}=\Sigma_{i}p_{ij}$).
According to the Bayes rule, the conditional probability can be written as $p_{i|j}=p_{ij}/q_{j}$. As a result, classical mutual information has the equivalent form
\begin{equation}
J(A:B)=H(A)-H(A|B), \label{conditional}
\end{equation}
where $H(A|B)$ is the conditional entropy of $A$ when $B$ is known, and $H(A|B)=-\Sigma_{i,j}p_{ij}\log_{2}p_{i|j}=H(A,B)-H(B)$.

In the quantum case, the density matrix $\rho$ fully characterizes the state of a system. The information of a certain state is given by the von Neumann entropy $\mathcal{S}(\rho)$, where $\mathcal{S}(\rho)=-\text{Tr}(\rho\log_{2}\rho)$. For a bipartite quantum system with subsystems $A$ and $B$, the quantum mutual information is represented as
\begin{equation}
\mathcal{I}(\rho_{AB})=\mathcal{S}(\rho_{A})+\mathcal{S}(\rho_{B})-\mathcal{S}(\rho_{AB}), \label{quantummutual}
\end{equation}
where $\rho_{A}=\text{Tr}_{B}\rho_{AB}$ ($\rho_{B}=\text{Tr}_{A}\rho_{AB}$) is the reduced density matrix of partition $A$ ($B$). Equation (\ref{quantummutual}) is a direct quantum extension of Eq. (\ref{mutual}). Because a measurement generally disturbs the quantum state, the conditional entropy of subsystem $A$ changes when we choose different measurements on subsystem $B$, and the extension of Eq. (\ref{conditional}) to the quantum region is not so direct.

Consider a set of measurements $\{M_{B}^{k}\}$ on subsystem $B$. The reduced state of subsystem $A$ conditioned on the measurement labeled by $k$ becomes
\begin{equation}
\rho_{A}^{k}=\frac{1}{r_{k}}\text{Tr}_{B}[(\mathbf{1}_{A}\otimes M_{B}^{k})\rho_{AB}(\mathbf{1}_{A}\otimes M_{B}^{k})^{\dag}],
\end{equation}
with probability $r_{k}=\text{Tr}_{AB}[(\mathbf{1}_{A}\otimes M_{B}^{k})\rho_{AB}(\mathbf{1}_{A}\otimes M_{B}^{k})^{\dag}]$ and with $\mathbf{1}_{A}$ representing the identical operator on the subsystem $A$. As a result, the conditional entropy of subsystem $A$ due to the measurement on $B$ can be defined as
\begin{equation}
\mathcal{S}(A|B)=\Sigma_{k}r_{k}\mathcal{S}(\rho_{A}^{k}). \label{eq:conditionalentropy}
\end{equation}
The quantum generalization of Eq. (\ref{conditional}) becomes
\begin{equation}
J(\rho_{AB})=\mathcal{S}(\rho_{A})-\mathcal{S}(A|B).
\end{equation}
and the maximum extractable information from the measurement on $B$
\begin{equation}
\mathcal{C}(\rho_{AB})=\max_{M_{B}^{k}}[J(\rho_{AB})]=\max_{M_{B}^{k}}[\mathcal{S}(\rho_{A})-\mathcal{S}(A|B)].  \label{classical}
\end{equation}
is then defined as the classical correlation of $\rho_{AB}$\cite{Henderson2001}.
The two generalized quantum mutual information sets are generally not equal to each other and their difference is defined as the famous quantum discord\cite{Ollivier2001}
\begin{equation}
\mathcal{Q}(\rho_{AB})=\mathcal{I}(\rho_{AB})-\mathcal{C}(\rho_{AB}). \label{discord}
\end{equation}
The quantum mutual information $\mathcal{I}(\rho_{AB})$ is equal to the sum of classical correlation $\mathcal{C}(\rho_{AB})$ and quantum discord $\mathcal{Q}(\rho_{AB})$, and which is typically used to quantify the total correlations for bipartite quantum systems\cite{Groisman2005,Schumacher2006}.

From Eq. (\ref{classical}) and Eq. (\ref{discord}), one can see that the one-sided quantum discord is not symmetric; when the measurement is performed on $A$, the result may be different. However, there are other measures of quantum and classical correlations available for combined quantum systems. A symmetrical method with two-sided measurement over both subsystems of a bipartite system is proposed to quantify the classical correlation represented by the maximal classical mutual information\cite{DiVincenzo2004,Terhal2002} and the similar symmetric quantification of nonclassical correlation is defined\cite{Piani2008,Wu2009}. A thermodynamic approach is also proposed\cite{Oppenheim2002,Horodecki2005}, where the quantum-information deficit--the difference between the total information and the extractable information using closed local operation and classical communication (CLOCC)--is used to quantify the quantum correlation\cite{Horodecki2005}. Classical and quantum correlations are characterized using measurement-induced disturbance by considering the property that classical states can be measured without disturbance \cite{Luo2008_2}. Modi {\it et al.} proposed using the relative entropy as a distance measure of correlations--in a fashion similar to the relative entropy of entanglement\cite{Vedral1997}--which provides a unified view of quantum and classical correlations\cite{Modi2010}. A similar geometric measure of quantum correlations was proposed based on the Hilbert-Schmidt distance measurement\cite{Dakic2010}. A geometric measurement-induced nonlocality is also defined\cite{Luosl2011}, which is further reformulated in an entropic way\cite{Huml2012-2}. There are also other kinds of nonclassical correlation measures, and a recent review can be found in Ref.~\refcite{Modi2012}. Usually these definitions of nonclassical correlation are not equal to each other. However, some of the measures are equivalent for the a special kind of Bell-diagonal state\cite{Modi2010,Maziero2011}.

\section{The system-environment dynamics of quantum discord}

\begin{figure}[tbph]
\begin{center}
\includegraphics [width= 3.0in]{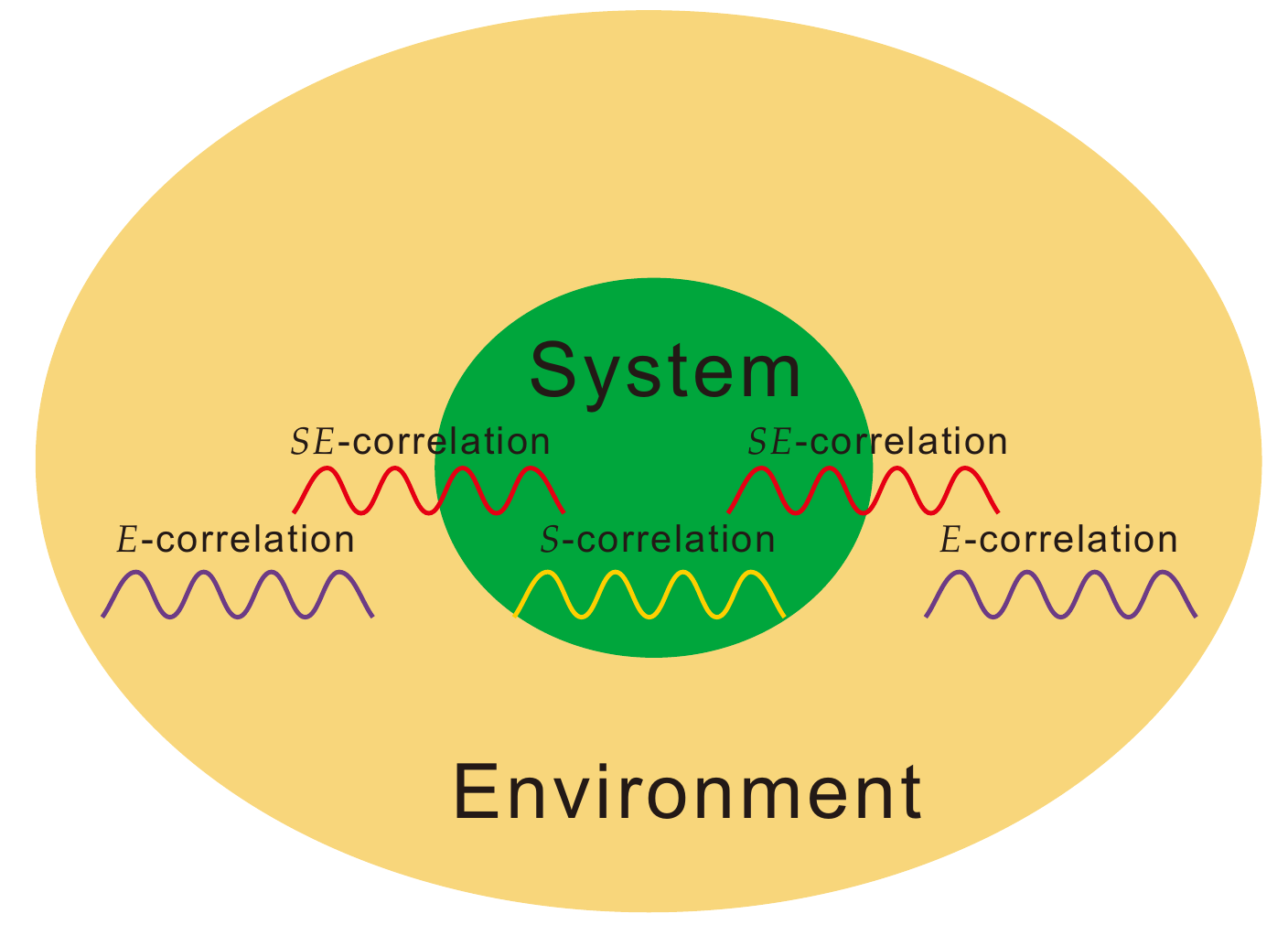}
\end{center}
\caption{Schematic of an open quantum system with its corresponding correlation dynamics (represented by the wavy lines).}\label{fig:coupling}
\end{figure}

As shown in fig. \ref{fig:coupling}, the open quantum system $S$ is coupled to its environment $E$, with the correlations evolving between $S$ and $E$ ($SE$-correlation) and within their own systems ($S$-correlation and $E$-correlation).  The total Hamiltonian of the combined system can be written as\cite{Breuer2002}
\begin{equation}
H=H_{S}\otimes \mathbf{1}_{E}+\mathbf{1}_{S}\otimes H_{E}+H_{int},
\end{equation}
where $H_{S}$ is the self-Hamiltonian of the open system $S$, $H_{E}$ is the free Hamiltonian of the environment $E$, $H_{int}$ is the interaction Hamiltonian between the system and the environment, and $\mathbf{1}$ represents the identity operator.

The evolution of the closed total system ($S+E$), is described by the unitary operation $U(t)$, where the final state becomes
\begin{equation}
\rho_{SE}(t)=U(t)\rho_{SE}(0)U(t)^{\dag}.
\end{equation}
The reduced density matrix of the open quantum system $S$ becomes $\rho_{S}=\text{Tr}_{E}(\rho_{SE})$, and the evolution of the quantum system can be expressed as a dynamic map
\begin{equation}
\rho_{S}(0)\rightarrow\rho_{S}(t)=\Phi(\rho_{S}(0))\equiv\text{Tr}_{E}\{U(t)\rho_{SE}U(t)^{\dag}\}. \label{eq:map1}
\end{equation}

Fragile correlations in open quantum systems are easily destroyed by unavoidable noises, so knowledge of the dynamic behavior of correlations helps in designing suitable protocols to protect useful resources during processing. Moreover, correlations can have distinctive dynamic behaviors, which can be of both practical and fundamental importance. One of the most distinctive dynamic properties of quantum entanglement is that it may suffer from sudden death, i.e., entanglement completely disappears at a finite time in the system's evolution\cite{Yu2009}. A quantum discord investigation of open quantum systems that naturally compares other dynamic behaviors to entanglement obtains several interesting results. First, there is not sudden death of quantum discord during its dynamic phase\cite{Werlang2009,Fanchini2010,Wang2010} and almost all states have nonclassical correlations\cite{Ferraro2010}. Secondly, during the dynamics of quantum discord and its classical correlation counterpart, there is a sudden change of behavior in their decay rates\cite{Maziero2009}. Both quantum and classical correlations are unaffected by the decoherence noise and the open system dynamics exhibits a sudden transition from the classical to the quantum decoherence regime\cite{Mazzola2010}. Thus, in this paper, we focus on the system-environment dynamics of quantum discord and the fundamental influence of initial system-environment states on the dynamics of open quantum systems.

As shown in fig. \ref{fig:coupling}, the interaction Hamiltonian between the system and environment leads to certain system-environment correlations that affect the states of the open quantum system. Maziero {\it et al.} studied the transference of correlation between the system and its reservoirs by considering a non-interacting two-qubit composite system ($AB$) under the influence of two independent environments ($E_{A}$ and $E_{B}$)\cite{Maziero2010}. By employing the Kraus operator-sum representation, the group analytically and numerically analyzed the dynamics of correlations of different partitions in different kinds of noise channels: amplitude damping, phasing damping, bit flip, bit-phase flip and phase flip. For the amplitude-damping channel--the dissipative interaction between the system and the environment that includes an exchange of energy--the action of the channel over one qubit is represented as
\begin{eqnarray}
|0\rangle_{S}\otimes|0\rangle_{E}&\rightarrow&|0\rangle_{S}\otimes|0\rangle_{E},\nonumber \\
|1\rangle_{S}\otimes|0\rangle_{E}&\rightarrow&\sqrt{1-p}|1\rangle_{S}\otimes|0\rangle_{E}+\sqrt{p}|0\rangle_{S}\otimes|1\rangle_{E}, \label{eq:amplitude}
\label{quantum map}
\end{eqnarray}
where $|0\rangle_{S}$ ($|1\rangle_{S}$) is the ground (excited) qubit state and $|0\rangle_{E}$ ($|1\rangle_{E}$) represents the state of the environment with zero (one) excitation distributed over all its modes. The initial state of the whole system can be written as
\begin{equation}
\rho_{ABE_{A}E_{B}}=\frac{1}{4}(\mathbf{1}^{AB}+\sum_{i=1}^{3}c_{i}\sigma_{i}^{A}\otimes\sigma_{i}^{B})\otimes|00\rangle_{E_{A}E_{B}},
\end{equation}
where $|00\rangle_{E_{A}E_{B}}$ is the vacuum state of the environments and $\sigma_{i}$ ($i=1,2,3$) are the three Pauli matrixes. The coefficients $c_{i}$ are real numbers and the state of the system $\rho_{AB}$ represents Bell diagonal states. The corresponding reduced-density operator is obtained by taking the partial trace of $\rho_{ABE_{A}E_{B}}$.

Maziero {\it et al.} considered the correlation dynamics of $AB$, $AE_{A}$, $AE_{B}$ and $E_{A}E_{B}$\cite{Maziero2010}. Note that, due to the symmetry of the system, the density matrix of the partition $BE_{B}$ and $BE_{A}$ are identical to that of $AE_{A}$ and $AE_{B}$, respectively. For a Werner initial state where $c_{1}=c_{2}=c_{3}=-\alpha$ ($0\leq\alpha\leq1$), both the classical and quantum correlations of $AB$ vanished in the asymptotic limit. But the entanglement suffers from sudden death at a given time evolution. This phenomenon of entanglement sudden death between the qubits, and entanglement sudden birth between the reservoirs, was also observed in Ref.~\refcite{Lopez2008}. The vanishing of classical and quantum correlations between the parts of system $AB$ was accompanied by the simultaneous creation of corresponding correlations between the reservoirs $E_{A}E_{B}$. Because the qubits were under the influence of the local environment, the quantum correlations were created between each qubit and its own reservoir, i.e. betwen $AE_{A}$ and $BE_{B}$.

The authors also investigated\cite{Maziero2010} the phase-damping channel case, which describes the loss of quantum coherence without loss of energy. In this case, the interaction map for one qubit case becomes
\begin{eqnarray}
|0\rangle_{S}\otimes|0\rangle_{E}&\rightarrow&|0\rangle_{S}\otimes|0\rangle_{E},\nonumber \\
|1\rangle_{S}\otimes|0\rangle_{E}&\rightarrow&\sqrt{1-p}|1\rangle_{S}\otimes|0\rangle_{E}+\sqrt{p}|1\rangle_{S}\otimes|1\rangle_{E}. \label{eq:phase}
\label{quantum map}
\end{eqnarray}
For an initial separable system state with an independent environment, there is no entanglement created between the system and its own environment; i.e., $AE_{A}$ and $BE_{B}$ are still separable. However, decoherence does occur for the asymptotic decay of the off-diagonal elements of the initial state. As a result, information transference between the system and the environment relies on both classical and quantum correlations. For the phase-damping channel case, quantum correlations disappear in all partitions, a result different from the amplitude-damping case, where the quantum correlations are completely transferred from system $AB$ to reservoir $E_{A}E_{B}$ at $p=1$. The effect of the bit-flip, bit-phase-flip, and phase-flip channels is similar to that of the phase-damping channel, which destroys the phase information without an exchange of energy. Therefore, the dynamic behavior of the correlations in the these three channels is the same as in the phase-damping channel\cite{Maziero2010}.

Ge {\it et al.} analyzed\cite{Ge2010} the case of two spins interacting independently with
their own boson reservoirs with the corresponding Hamiltonian given by
\begin{equation}
H=\sum_{i=1}^{2}(\frac{\omega}{2}\sigma_{i}^{z}+\sum_{k=1}^{N}\omega_{k}b_{i,k}^{\dag}b_{i,k})
+\sum_{i=1}^{2}\sum_{k=1}^{N}g_{i,k}(\sigma_{i}^{-}b_{i,k}^{\dag}+\sigma_{i}^{+}b_{i,k}), \label{eq:spinboson}
\end{equation}
Here, $N\rightarrow\infty$. $b_{i,k}$ ($i=1,2$) is an
annihilation operator of the $k$-th model in the $i$-th reservoir with the
corresponding frequency $\omega_{k}$. $\sigma_{i}^{z}$,
$\sigma_{i}^{+}$, ($\sigma_{i}^{-}$) and $\omega$ represent the
Pauli operator, raising (lowering) operator, and Zeeman splitting of
the $i$-th spin, and $g_{i,k}$ denotes the coupling strength between
the $k$-th model in the $i$-th reservoir and the corresponding spin.
The group first considered an initial state with two excitations in the spin system
\begin{equation}
|\Psi_{0}\rangle=(\alpha|0\rangle_{s_{1}}|0\rangle_{s_{2}}+\beta|1\rangle_{s_{1}}|1\rangle_{s_{2}})|0\rangle_{r_{1}}|0\rangle_{r_{2}},
\end{equation}
with $|\alpha|^{2}+|\beta|^{2}=1$. The collective state $|0\rangle_{r_{i}}=\prod_{k=1}^{N}|0_{k}\rangle_{r_{i}}$ indicated there was no excitation in reservoir $r_{i}$. By introducing the collective state $|1\rangle_{r_{i}}$ representing the case of one excitation in the reservoir, they obtained the reduced density matrix of $\rho_{s_{1}s_{2}}$, $\rho_{s_{1}r_{1}}$, $\rho_{s_{2}r_{2}}$ and $\rho_{r_{1}r_{2}}$. The authors then looked at two kinds of spectral distribution. For a flat spectral density, both quantum and classical correlations initially stored in the spin system decayed monotonously, only gradually transferring to the reservoirs. On the other hand, the Lorentz form of spectral density resulted in a strong non-Markovion effect. The group also found correlations between all bipartitions of the system and reservoir oscillations, and that both the quantum and classical correlations stored in the spin system also transferred to the reservoirs over time. During the evolution of this kind of initial input state, quantum and classical correlations are equal to one another in the two-spin system and the two reservoirs.

In this work, another state with only one excitation in the spin system is further considered\cite{Ge2010}, i.e.
\begin{equation}
|\Psi_{0}\rangle=(\alpha|0\rangle_{s_{1}}|1\rangle_{s_{2}}+\beta|1\rangle_{s_{1}}|0\rangle_{s_{2}})|0\rangle_{r_{1}}|0\rangle_{r_{2}}.
\end{equation}
For this case, quantum correlation is no longer equal to classical correlation in the spin system or the reservoirs. Both the quantum and classical correlations initially stored in the spin system transfer to the reservoirs at the end of the system's evolution.

The transference of quantum correlation between systems and reservoirs has also been considered in a multi-qubit systems. Man {\it et al.}\cite{Man2011} studied a simple system of three entangled qubits ($A$, $B$, and $C$) which were coupled to three independent reservoirs ($a$, $b$, and $c$), and investigated the time evolution of discord for any two qubits. The Hamiltonian for this total system can be expressed as the sum of three independent spin-boson models
\begin{equation}
H=H_{Aa}+H_{Bb}+H_{Cc}
\end{equation}
where $H_{Xx}=\omega_{q}\sigma_{X}^{+}\sigma_{X}^{-}+\sum_{j}\omega_{j}x_{j}^{\dag}x_{j}+\sum_{j}(g_{j}\sigma_{X}^{+}x_{j}+
g_{j}^{*}\sigma_{X}^{-}x_{j}^{+})$, $Xx\in\{Aa,Bb,Cc\}$, which is similar to the Hamiltonian of Eq. (\ref{eq:spinboson}). The qubits $A$, $B$, and $C$ are initially prepared in the extended W-like state and read as
\begin{equation}
\rho_{ABC}(0)=\frac{1-p}{8}\mathbf{1}_{ABC}+p|W(0)\rangle_{ABC}\langle W(0)|,
\end{equation}
where $|W(0)\rangle_{ABC}=\alpha|001\rangle+\beta|010\rangle+\gamma|100\rangle$ with $|\alpha|^{2}+|\beta|^{2}+|\gamma|^{2}=1$, and $\mathbf{1}_{ABC}$ is the identity matrix. The three reservoirs $a$, $b$, and $c$ are initially in the vacuum state $|000\rangle_{abc}$, and the initial total system is in the product state
$\rho_{ABCabc}(0)=\rho_{ABC}(0)\otimes|000\rangle_{abc}\langle000|$.
The reduced matrix operator of any two qubits, e.g. qubits $AB$, is obtained by tracing the total evolved state $\rho_{ABCabc}(t)$ over the reservoirs $abc$ and the qubit $C$; the other reduced density matrixes can be obtained in a similar fashion.  Man {\it et al.}\cite{Man2011} first considered a Markovian evolution case with $\alpha=\beta=\gamma=1/\sqrt{3}$. When $p=1$ (W-state), the group found that both the entanglement and discord of qubits $AB$ decayed monotonously to zero and transferred to the corresponding reservoirs $ab$. On the other hand, for $p=0.8$ (mixed state) and $p=0.5$ (separated state), the quantum discord initially in the qubits $AB$ transferred to the reservoirs $ab$ but with a sudden change in the behaviors of their dynamics rates. The group showed that while entanglement can suffer from sudden death and sudden birth, quantum discord always evolves asymptotically.
The authors further considered\cite{Man2011} a model with two remote nodes, each of which consisted of two qubits coupled to a common reservoir, i.e., qubits $AC$ in reservoir $a$ and qubits $BD$ in $b$. The initial state of the total system was prepared to be $\rho_{ACaBDb}(0)=\rho_{AB}(0)\otimes\rho_{CD}(0)\otimes\rho_{ab}(0)$, where $\rho_{AB}=(1-p)/4\mathbf{1}_{AB}+p|\Psi\rangle_{AB}\langle\Psi|$ with $|\Psi\rangle_{AB}=\alpha|00\rangle+\beta|11\rangle$, $\rho_{CD}=|00\rangle_{CD}\langle00|$, and $\rho_{ab}=|00\rangle_{ab}\langle00|$. As a result, the two nodes were initially correlated by the correlations between the qubits $A$ and $B$. Both Markovian and non-Markovian cases were investigated, with the group finding that the quantum discord of $AB$, $CD$, and $ab$ all tended towards long-term steady values, while the discord of $ab$ was always larger than that of $AB$ and $CD$, and the entanglement of $AB$ and $CD$ eventually decayed to zero. Similar to the first model, there were sudden changes in behavior in the quantum discord of the qubits and the reservoirs when the  initial states of $AB$ were prepared in a mixed system.

In the above discussion, the environments were modeled as the quantum system. However, what is the result when the environment is not quantum but classical? Lo Franco {\it et al.} considered\cite{Franco2012} a pair of noninteracting qubits, each locally
coupled to a classical random external field. The external field was equal for both qubits and was unaffected by the coupled qubit. The amplitude of the external field was fixed, but
the phase of each mode was classically random and equal to either 0 or $\pi$ with a probability of $1/2$ (the case for which $p\neq1/2$ was also considered by the authors). For each external field with phase 0 or $\pi$, the qubits underwent a unitary evolution. The dynamical map for the initial two qubits $\rho_{AB}$ can
be written as (when $p=1/2$)
\begin{equation}
\rho_{AB}(t)=\frac{1}{4}\sum_{i,j=1}^{2}U_{i}^{A}(t)U_{j}^{B}\rho_{AB}(0)U_{i}^{A\dag}U_{j}^{B\dag},
\end{equation}
with $U_{j}^{S}(t)=e^{-iH_{j}t/\hbar}$. Here, ($S=A,B$) represents the time
evolution operator, where $H_{j}=i\hbar
g(\sigma_{+}e^{-i\phi_{j}}-\sigma_{-}e^{i\phi_{j}})$. The time
evolution operator $U_{j}^{S}(t)$ can be expressed in the basis $\{|0\rangle,|1\rangle\}$ as
\begin{equation}
U_{j}^{S}(t)=\left(
\begin{array}{cc}
\cos(gt) & e^{-i\phi_{j}}\sin(gt) \\
-e^{i\phi_{j}}\sin(gt) & \cos(gt)%
\end{array}%
\right),\label{X}
\end{equation}
where $j=1,2$ with $\phi_{1}=0$ and $\phi_{2}=\pi$.

When the initial state was prepared in the Bell diagonal state of
$\rho_{AB}=\lambda_{+}|\psi_{+}\rangle\langle\psi_{+}|+\lambda_{-}|\psi_{-}\rangle\langle\psi_{-}|$,
with $\lambda_{+}=0.9$ and $\lambda_{-}=0.1$ ($|\psi_{\pm}\rangle=1/\sqrt{2}(|01\rangle\pm|10\rangle)$), the authors observed
the collapse and revival of entanglement,
quantum discord, and classical correlation, with all the correlations
measured in the unified relative entropy form\cite{Modi2010}. The group determined that classical correlation oscillates when quantum discord remains constant, and vice versa. Thus, classical correlation and quantum discord are equal to each other at certain times in the system's evolution, which is the same as the case of a sudden transition between classical and quantum decoherence\cite{Mazzola2010}.

It is surprising that the classical environment has no back actions and does
not store the quantum information of the two qubits. Tasking into account the
correlations in the quantum-classical state of the system-reservoir
system, the authors suggested that the recorded operations applied to the
qubits in the classical environment could play an important role in
retrieving the quantum correlations\cite{Franco2012}.
The group also discussed a
class of global system-environment evolutions possessing a lack of back-action. For this class, any reduced system dynamics obtained with a
quantum environment can also be obtained by modeling the classical environment\cite{Franco2012}.
In order to interpret the revival of correlations, the authors generalized the non-Markovian quantifier introduced in Ref.~\refcite{Rivas2010}
by using a time-dependent non-Markovianity quantifier
\begin{equation}
I^{Rn}(t)=\int_{t_{0}}^{t}|\frac{dRn[\rho_{AB}(t')]}{dt'}|dt'-\triangle
Rn(t),
\end{equation}
Here, $Rn$ represents the relative entropy of entanglement\cite{Vedral1997}, with $\triangle
Rn(t)=Rn(t)-Rn(t_{0})$ where $t_{0}\leq t$. The authors observed that $I^{Rn}(t)$
monotonically increased in the same time regions when revivals of
entanglement and quantum discord occurred\cite{Franco2012}.

The initial correlations between the system and environment can fundamentally
influence the dynamics of the system. One of the most famous examples of this is the
completely positive map for describing the dynamics of open quantum
systems\cite{Breuer2002}. If the dynamical map Eq. (\ref{eq:map1}) can be expressed as
\begin{equation}
\rho_{S}(t)=\Phi(\rho_{S}(0))=\sum_{i}\alpha_{i}C_{i}\rho_{S}(0)C_{i}^{\dag}, \alpha_{i}\geq0 \; \forall\; i \label{eq:sumrepresentation}
\end{equation}
where $C_{i}$ is the corresponding matrices and $\sum_{i}C_{i}C_{i}^{\dag}=\mathbf{1}$, it is a completely positive map\cite{Choi1975}, and thus can described the dynamics of the system when the initial
state between the system and its environment is a product
state\cite{Sudarshan2003}. In a recent work, Rodr\'{i}guez {\it et
al.} extended\cite{Rodriguez2008} this result and demonstrated that, if
the initial state between the system and environment is a
classically correlated state with zero discord, the dynamics of the open quantum system are
a completely positive map. The system-environment state with vanishing quantum discord can be represented as $\rho_{SE}=\sum_{i}p_{i}M_{i}^{S}\otimes\rho_{E|i}$, where $\rho_{E|i}$ are density matrices for the environment and $\{M_{i}^{S}\}$ is a complete set of orthogonal projectors on the system $S$ with $p_{i}\geq0$ and $\sum_{i}p_{i}=1$. For such initial system-environment states, Rodr\'{i}guez {\it et
al.} proved\cite{Rodriguez2008} that, under the unitary evolution of the total systems, the final state of the system can be expressed in the form of Eq. (\ref{eq:sumrepresentation}), which implies a completely dynamical map.
By introducing a kind of special-linear state with the property of being of full measure in the set of mixed bipartite states, Shabani and Lidar demonstrated\cite{Shabani2009} that a quantum dynamical process with the form of Eq. (\ref{eq:map1}) is always a linear Hermitian map for arbitrary initial system-environment states ($\rho_{SE}(0)$). They further proved that the vanishing quantum discord of $\rho_{SE}(0)$ is a necessary and sufficient condition to induce a completely positive map.

Coupling between the system and the environment leads to decoherence and changes the entropy of the system, and von Neumann entropy $\mathcal{S}$ is typically used to quantify this decoherence. Thus, the rate of decoherence is related to the entropy rate of the system, which is defined as
$\frac{d\mathcal{S}}{dt}$.
Recently, a general result for determining a system's entropy rate has been demonstrated\cite{Rodriguez2011}. The sufficient and necessary condition for a zero system entropy rate at a time $\tau$ under any system-reservoir coupling is expressed as
\begin{equation}
\left[ \frac{d}{dt}\mathcal{S}\right]_{t=\tau} = 0\;\;\forall\; H \quad \Leftrightarrow \quad \left[\rho^{S}\otimes \mathbf{1}^{E},\rho^{SE}\right]=0.
\end{equation}
A system-environment state with the property of $\left[\rho^{S}\otimes \mathbf{1}^{E},\rho^{SE}\right]=0$ is called a lazy state, which has the form of
$\rho^{SE} = \sum_{j} M_j^{S} \otimes \mathbf{1}^{E}  \rho^{SE} M_j^{S} \otimes \mathbf{1}^{E}$.
Note that $\rho^{S} = \sum_j p_j M_j^{S}$ and $\{M_j^{S}\}$ are orthonormal projectors. As a result, the lazy state is a classical state with vanishing quantum discord\cite{Rodriguez2011}.
The authors further provided a universal bound on the rate of decoherence for a system-environment interaction $H_{int}$ of arbitrary strength, which is expressed as
\begin{equation}
\biggl\lvert { \;\frac{d}{dt}\mathcal{S} \;\biggr\rvert_{t=\tau}} \le \bigl\lVert H_{int}\bigr\rVert \; \bigl\lVert \left[\ln(\rho^{S})\otimes \mathbf{1}^{E},\;\rho^{SE}\right]\bigr \rVert _1.
\end{equation}

Due to the fact that almost all quantum states have nonclassical
correlations\cite{Ferraro2010}, one may conclude that the rate of
system entropy will be always high. However, in another recent
work\cite{Hutter2012}, the authors demonstrated that this is not the
case. The authors showed that the probability for obtaining a high entropy rate for a randomly chosen state
$\rho_{SE}$ is very small, i.e.
\begin{equation}
\mathrm{Pr}_{\rho_{SE}}\left[| \frac{d}{dt}\mathcal{S}\bigr| \geq \bigl\lVert H_{int}\bigr\rVert_{\infty}\varepsilon\right]\leq\delta,
\end{equation}
where $\varepsilon=2^{-1/2(\log_{2} d_{E}-3\log_{2} d_{S}-4)}$ and
$\delta=2e^{-d_{S}^{2}/16}$ with $d_{S}$ and $d_{E}$ representing the dimensions of the system ($S$) and the environment ($E$), respectively. For a sufficiently large environment ($\log_{2}
d_{E}>3\log_{2} d_{S}$) and not too small of a system ($\log_{2}
d_{S}>2$), the authors found that the entropy rate of $S$ in a vast majority of bipartite
states can only decay at a
vanishing rate. For small systems where $\log_{2} d_{S}\leq2$, a similar
bound exists when $\log_{2} d_{E}>(9/2)\log_{2} d_{S}$\cite{Hutter2012}.

\section{External control effects on open quantum systems}

One can greatly influence the dynamics of correlations by implementing operational control on an open quantum systems. In the work of
Francica {\it et al.}, the group considered\cite{Francica2010} quantum
Zeno and anti-Zeno effects on quantum and classical correlations of
two noninteracting qubits coupled to a
commonly structured reservoir, e.g. two atoms coupled to a common zero-temperature bosonic reservoir.
The authors performed nonselective measurements on the
system $AB$ at time interval $T$  to observe the quantum Zeno effect on entanglement, and found two characteristics: 1) one of the possible measurement
outcomes is a projection onto the collective ground state $\vert
\psi_{0} \rangle_{S} = \vert 0 \rangle_{A} \vert 0 \rangle_{B}$, and 2) the
measurement cannot distinguish between the excited-states
$|\psi_{1}\rangle_{S}=|1\rangle_{A} |0\rangle_{B}$ and
$|\psi_{2}\rangle_{S}=|0\rangle_{A}|1\rangle_{B}$\cite{Maniscalco2008}.

The measurements can thus be described by the two following projectors
\begin{eqnarray}
  M_0 &=& |\psi_{0}\rangle_{S}\langle\psi_{0}|\otimes
    \mathbf{1}_E, \nonumber\\
  M_1 &=& \left( |\psi_{1}\rangle_{S}\langle\psi_{1}|+|\psi_{2}\rangle_{S}\langle|\psi_{2}|\right)\otimes
    \mathbf{1}_E, \label{Pi1}
\end{eqnarray}
In this way, the group separate results for the different cases of off-resonant interaction between the qubits and the
cavity mode and the resonant case\cite{Maniscalco2008}. The authors observed that, under certain conditions, the same
measurements that greatly suppressed entanglement
loss could instead lead to a much faster disentanglement, showing the anti-Zeno effect\cite{Kofman2000,Facchi2001}. the group further found that the dynamics of quantum discord and classical correlation are qualitatively similar to the dynamics of entanglement, and a series of oscillations between the Zeno and
anti-Zeno effects can occur as a function of the time delay between
successive measurements. Studying the fact that classical correlation can be
protected by a sufficiently high measurement frequency that can also display the Zeno
and anti-Zeno oscillation could lead to a deeper understanding of
classicality.

Bang-bang control technology, as a dynamical decoupling strategy, has been used to prevent quantum decoherence\cite{Viola1998}. Xu {\it et al.} applied the bang-bang control pulses to two
noninteracting two-level atoms in an independent environment to investigate its influences on the dynamics of quantum correlations\cite{Xu2011_2}.
When adding the bang-bang pulses, the Hamiltonian of one atom interacting with its own reservoir can be expressed as
\begin{equation}
H=H_{0}+H_{int}+H_{p},
\end{equation}
with $H_{0}=\frac{\omega_{0}}{2}\sigma_{z}+\omega a^{\dag}a$ and $H_{int}=g(\sigma_{+}a+\sigma_{-}a^{\dag})$, where $a$ and $a^{\dag}$ denote the annihilation and creation operators for the cavity field. The Hamiltonian for a train of identical pulses of duration $\tau$ is $H_{p}$, expressed as\cite{Xu2011_2}
\begin{equation}
H_{p}=V\sigma_{z}\sum_{n=0}^{\infty}\theta(t-T-n(T+\tau))\theta((n+1)(T+\tau)-t),
\end{equation}
where $V$ is the amplitude of the control field and $T$ is the time interval between two consecutive pulses. In the authors' case, $V$ was set to $\pi/2\tau$, which implies the $\pi$-pulse only.
The two atoms were initially prepared to be the Werner-like states
\begin{equation}
\rho_{AB}=x|\Phi\rangle\langle\Phi|+\frac{1-x}{4}I,
\end{equation}
where $|\Phi\rangle=\alpha|0\rangle_{A}|1\rangle_{B}+\beta|1\rangle_{A}|0\rangle_{B}$ with $|\alpha|^{2}+|\beta|^{2}=1$ and $x$ represents the purity of the initial states. The cavity field was prepared in the Fock state or the thermal state. The group determined that both entanglement and quantum discord can be enhanced by the application of bang-bang control, and the increased amount is larger with shorter time intervals ($T$) between the control pulses\cite{Xu2011_2}. When the cavity field was initially in the vacuum state, at time $t_{2N}=2N(T+\tau)$, quantum discord was found to recover to the initial value for the detuning $\delta=0$ ($\delta=\omega_{0}-\omega$). When $\delta\neq0$, the quantum discord fluctuated at time $t_{2N}$ with period $2\pi/\delta$ and the oscillate amplitude was smaller with the shorter interval. The maximal revivals of quantum discord in the thermal state case decreased slightly with time, and the decrease became even slower with shorter control pulses.

Li {\it et al.} investigated\cite{Li2011_2} the effects of practical feedback control
on the dynamics of entanglement and geometric quantum discord\cite{Dakic2010} of two identical atoms ($A$ and $B$) resonantly coupled to a single-mode cavity by
considering the Markovian feedback\cite{Wiseman1993}. In this system, the control
Hamiltonian is $H_{c} = I(t) F$, where $I(t)$
is the signal outputted by the homodyne detection and $F$ is the feedback Hamiltonian.
Taking the efficiency of the detection, denoted by $\eta$, into consideration, the modified master equation for the state of the two atoms becomes\cite{Yamamoto2005}
\begin{equation}
\frac{{d\rho }}{{dt}} =  - i[H + \frac{1}{2}(G^\dag  F + FG),\rho ] + D[G - iF]\rho  + D[\sqrt {\frac{{1 - \eta }}{\eta }} F]\rho.
\end{equation}
where $H = \Omega
(\sigma _x^{(A)}  + \sigma _x^{(B)} )$ represents the driving of the laser and $D[X]\rho  = X\rho X^\dag   - (X^\dag  X\rho  + \rho
X^\dag  X)/2$ represents evolution due to
coupling with the operator $X$ between the system and the environment. Note that $G=\Gamma(\sigma_{-}^{A}+\sigma_{-}^{B})$, which represents the jump operator ($\Gamma$ was set to be 1 in this work). If both atoms are initially in the excited state ($|11\rangle_{AB}$), entanglement first increases and then decreases with feedback control. The phenomenon of entanglement sudden death occurs when $\eta<1$, and the duration of the non-entanglement period increases as $\eta$ decreases. Geometric quantum discord displays a similar tend to entanglement, but without sudden death. When both atoms are initially in the ground state ($|00\rangle_{AB}$), the noise of the detector ($\eta<1$) can trigger the detector and create entanglement and quantum discord between the two atoms. For the initial entangled input state $(|11\rangle_{AB}-|00\rangle_{AB})/\sqrt{2}$, the authors found that there was also a sudden change of entanglement with the perfect detection due to the discontinuous maximum value in calculating concurrence\cite{Wootters1998}.

\section{Experimental investigation of the dynamics of quantum discord in two-qubit systems}

Several distinctive properties of the dynamics of quantum and classical correlations under decoherence were experimentally investigated by Xu {\it et al.}\cite{Xu2010_3} using a phase-damping channel simulated by birefringent quartz plates as the noisy environment. The interaction quantum map was similar to that of Eq. (\ref{eq:phase}).
The information carriers were encoded as photon polarizations with the coupling between the photon's polarization and frequency modes occurring in the birefringent quartz plates. This coupling led to a phase-damping effect through tracing the frequency freedoms. By constructing unbalanced Mach-Zehnder devices, the group prepared different kinds of Bell-diagonal states, i.e. $\rho_{AB}=c_{1}|\Phi^{+}\rangle\langle\Phi^{+}|+c_{2}|\Phi^{-}\rangle\langle\Phi^{-}|
+c_{3}|\Psi^{+}\rangle\langle\Psi^{+}|+c_{4}|\Psi^{-}\rangle\langle\Psi^{-}|$ with $|\Phi^{\pm}\rangle=1/\sqrt{2}(|00\rangle\pm|11\rangle)$ and $|\Psi^{\pm}\rangle=1/\sqrt{2}(|01\rangle\pm|10\rangle)$. The minimum conditional entropy of Eq. (\ref{eq:conditionalentropy}) was obtained by scanning the measurement on photon $B$. Quantum discord and classical correlation were obtained from the final states reconstructed by quantum state tomography\cite{James2001}. For an initial state with $c_{2}=3/4$ and $c_{4}=1/4$ ($c_{1}$=0 and $c_{3}$=0), there is a sudden transition from the classical decoherence to the quantum decoherence regime\cite{Mazzola2010}, as shown in figure 2 (the transition point is represented by the crossover of the four panes with different colors). The figure also shows the non-entanglement quantum correlation, which exhibits a sudden change in the decay rate\cite{Maziero2009}. Quantum discord monotonically decays after the sudden transition point, while entanglement suffers from sudden death after a certain period. The experimental results are in good agreement with theoretical predictions. For a different initial state with $c_{1}=0.09, c_{2}=0.09, c_{3}=0.81$ and $c_{4}=0.01$, quantum correlation is larger than classical correlation as the system evolves, disproving an early conjecture that classical correlation is always larger than quantum correlation\cite{Lindblad1991}.

\begin{figure}[tbph]
\begin{center}
\includegraphics [width= 3.0in]{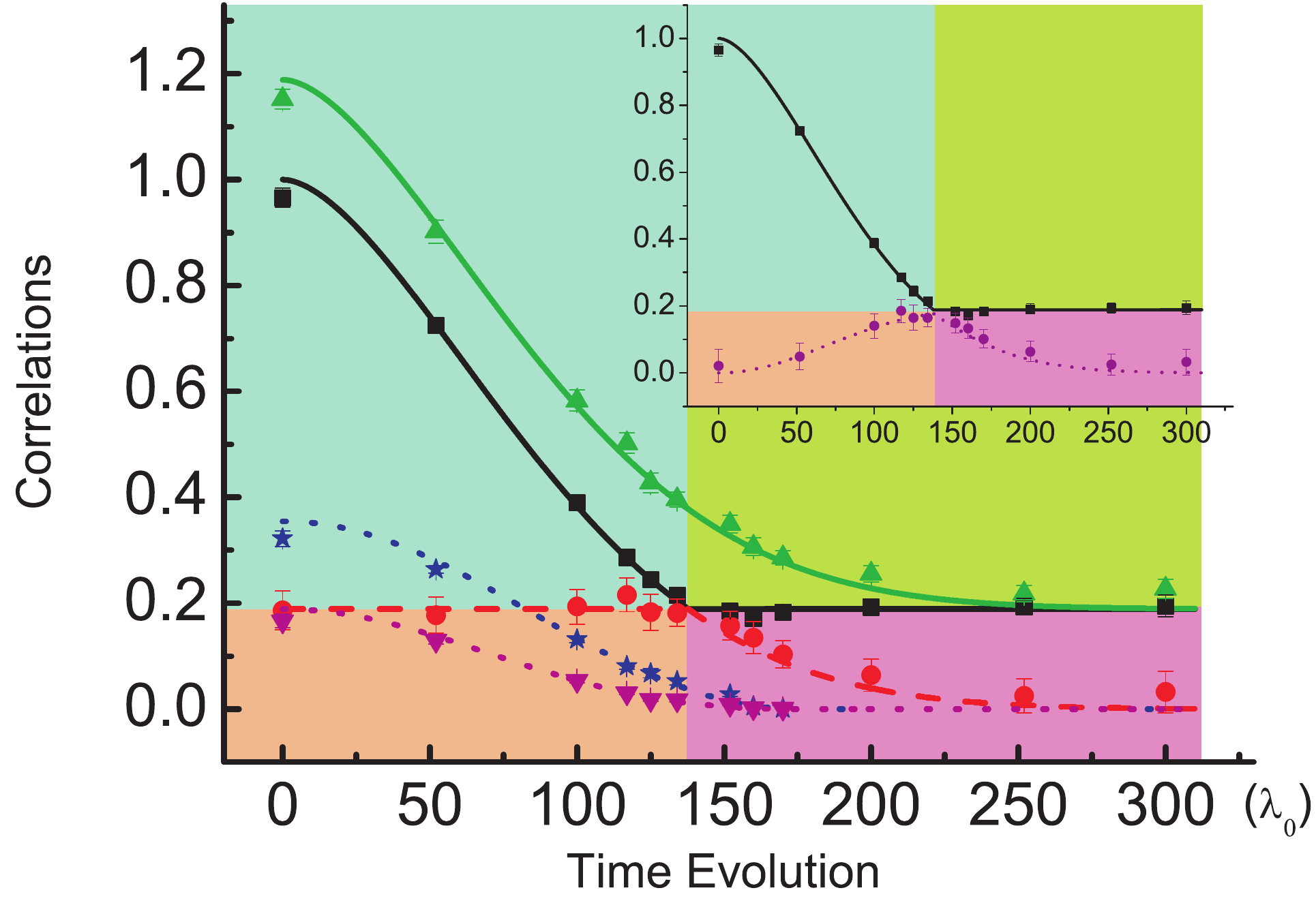}
\end{center}
\caption{The correlation dynamics of the input state
$\rho_{AB}=\frac{3}{4}|\Phi^-\rangle\langle\Phi^-|+\frac{1}{4}|\Psi^-\rangle\langle\Psi^-|$. The green upward-pointing triangles, black squares, red
dots, blue stars, and magenta downward-pointing triangles represent the
experimental results of $\mathcal{I}$ (mutual quantum information), $\mathcal{C}$ (classical correlation), $\mathcal{Q}$ (quantum correlation),
$En$ (entanglement of formation) and $Rn$ (relative entropy of entanglement), respectively, with the green solid line, black solid line, red
dashed line, blue dotted line and magenta dotted line representing
the corresponding respective theoretical predictions. Non-entanglement quantum
correlation ($\mathcal{D}$) is compared to $\mathcal{C}$
in the inset;
purple dots represent the experimental results of $\mathcal{D}$ and
the purple dotted line is the corresponding theoretical prediction.
The x-axis represents the total thickness of the quartz plates, with $\lambda_{0}=0.78$
$\mu$m corresponding to the time evolution of the photons. This figure is reproduced from Xu {\it et al.} Nat. Comm. 1, 7 (2010).}
\end{figure}

The non-Markovian dynamics of classical and quantum correlations was further investigated in an all-optical setup by Xu {\it et al.}\cite{Xu2010_4}. In their experiment, the group simulated a non-Markovian channel using a Fabry-Perot cavity followed by quartz plates acting on one of the two photons. The Fabry-Perot cavity filtered the frequency distribution of the photon, leading to a refocusing effect of the relative phase in the dephasing environment and inducing the non-Markovian effect. The sudden transition from the classical to the quantum decoherence regime occurred at the beginning of the evolution, similar to the case in fig. 2. Under such a non-Markovian environment, the quantum correlation (measured by the relative entropy of discord\cite{Modi2010} and calculated from the reconstructed density matrixes) revived from near zero before decaying again in the subsequent evolution. However, there was not revival of the classical correlation, which remained constant over the same period. This phenomenon is explained by the weak non-Markovian effect. The authors theoretically showed that both classical and quantum correlations are revived with a narrower frequency width (i.e. stronger non-Markovian effect). The group further implemented an $\sigma_{x}$ operation on the photon under decoherence and investigated the corresponding correlation dynamics. This process found a sudden transition from the quantum to the classical revival regime, as well as correlation echoes.

Experimental investigations of quantum discord and classical correlation under decoherence have also been implemented using NMR (nuclear magnetic resonance) systems. Because of the high-temperature expansion, a typical NMR-system density matrix for the effect two-qubit case can be written as $\rho=\frac{1}{4}\mathbf{1}+\epsilon\Delta\rho$, where $\mathbf{1}$ is the identity matrix, $\epsilon$ is the thermal energy, and $\Delta\rho$ represents the deviation density matrix\cite{Oliveira2007} that can be reconstructed by quantum state tomography. Under the action of a global environment, Soares-Pinto {\it et al.}\cite{Soares-Pinto2010} investigated the dynamics of correlations in a NMR quadrupolar system. The symmetric classical and quantum correlations\cite{DiVincenzo2004,Terhal2002}, which were computed from an experimentally reconstructed deviation-density matrix, were found to both decay exponentially. These experimental results agree well with the theoretical predictions. The authors further showed that, depending on the initial states, the classical correlation could be larger or smaller than the quantum correlation during the system evolution.

The sudden change in behavior in correlation dynamics has been further demonstrated in the NMR systems at room temperature\cite{Auccaise2011}. After preparing different kinds of initial Bell-diagonal states with specific relations between their components, the authors left the systems to evolve freely under decoherence. The environment of such an NMR system can be modeled by two independent quantum channels: the phase-damping and amplitude-damping channels. The authors observed a sudden change phenomenon for quantum discord and its classical counterpart due to the presence of phase-damping noise. On the other hand, the amplitude-damping noise led to a small decay of the classical correlation compared to a constant classical correlation after the sudden change point. Note also that the two nuclear spins finally relaxed to the Gibbs states.

\begin{figure}[tbph]
\begin{center}
\includegraphics [width= 3.0in]{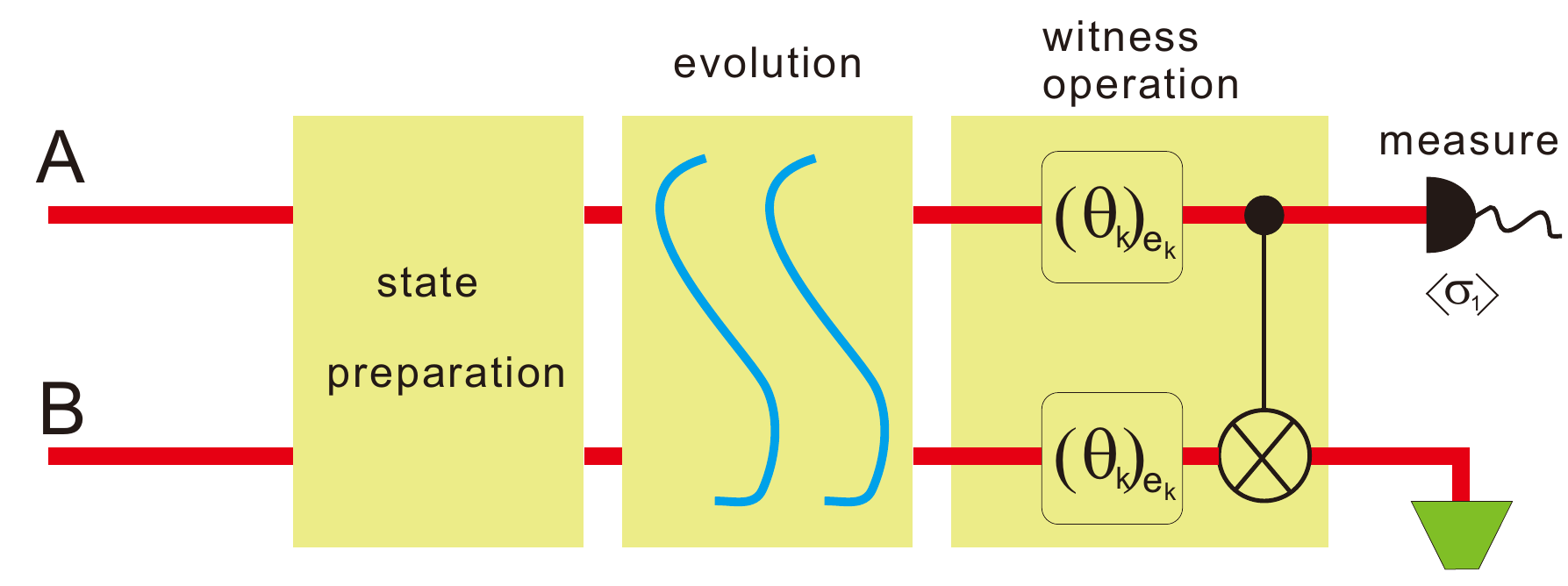}
\end{center}
\caption{Schematic setup for witnessing nonclassical correlations in a two-qubit system $AB$ under decoherence. This figure is reproduced from Auccaise {\it et al.} Phys. Rev. Lett. 107, 070501 (2011).} \label{fig:witness}
\end{figure}

Experimental investigation of quantum and classical correlations usually requires full quantum state tomography and their resulting numerical calculations. However, the concept of witnessing the quantumness of correlations recently received a great deal of attention. For a two-qubit states with the form of $\rho_{AB}=(\mathbf{1}^{AB}+\sum_{i=1}^{3}(a_{i}\sigma_{i}^{A}\otimes\mathbf{1}^{B}+b_{i}\mathbf{1}^{A}\otimes\sigma_{i}^{B}
+c_{i}\sigma_{i}^{A}\otimes\sigma_{i}^{B}))/4$, Maziero and Serra\cite{Maziero2010_4} obtained a sufficient condition for the states to be only classically correlated
\begin{equation}
\mathcal{W}_{\rho}=\sum_{i=1}^{3}\sum_{j=i+1}^{4}|\langle O_{i}\rangle_{\rho}\langle O_{j}\rangle_{\rho}|=0, \label{eq:witness}
\end{equation}
In this equation, $O_{i}=\sigma_{i}^{A}\otimes\sigma_{i}^{B}$, with $i=1,2,3$, and $O_{4}=\vec{z}.\vec{\sigma}^{A}\otimes\mathbf{1}^{B}+\mathbf{1}^{A}\otimes\vec{w}.
\vec{\sigma}^{B}$ with $\vec{z},\vec{w}\in\Re^{3}$ and $\vec{\sigma}\in\{\sigma_{1},\sigma_{2},\sigma_{3}\}$. Note that Eq. (\ref{eq:witness}) is also shown to be a necessary condition for the Bell-diagonal states. The value of $\langle \sigma_{i}^{A}\otimes\sigma_{i}^{B}\rangle$ ($\langle O_{i}\rangle$) can be changed to $\langle\sigma_{1}^{A}\otimes\mathbf{1}^{B}\rangle_{\gamma_{i}}$ with
$\gamma_{i}=U_{A\rightarrow B}[R_{e_{i}}(\theta_{i})\rho R_{e_{i}}^{\dag}(\theta_{i}))]U_{A\rightarrow B}$. In this equation, $R_{e_{i}}(\theta_{i})=R_{e_{i}}^{A}(\theta_{i})\otimes R_{e_{i}}^{B}(\theta_{i})$ with $R_{e_{i}}^{A(B)}(\theta_{i})$ representing a local rotation with angle $\theta_{i}$ around direction $e_{i}$ on qubit $A$ ($B$), $\theta_{1}=0$, $\theta_{2}=\theta_{3}=\pi/2$, $e_{2}=y$, $e_{3}=z$, and $U_{A\rightarrow B}$ represents the controlled-NOT gate between qubits $A$ and $B$ with $A$ as the control qubit.
Auccaise {\it et al.} experimentally demonstrated\cite{Auccaise2011_2} such a witness in a NMR system at room temperature by preparing different kinds of Bell-diagonal states (in this case $\langle O_{4}\rangle=0$) and performing a corresponding witness circuit, and the schematic setup is shown in fig. \ref{fig:witness} (where the evolution process is added in the case considering the witness dynamics). For each initial state, the experimental process was carried out three times to obtain the values $\langle\sigma_{1}^{A}\rangle_{\gamma_{i}}$ for determining the witness in Eq. (\ref{eq:witness}). Quantum state tomography and numerical optimization were used to calculate correlation values for comparison, and they agreed well with the witness results. The authors further investigated the decoherence dynamics of the witness with a freely evolving initial state. The witness tended to be 0, with the two-spin state evolving to a state with only classical correlation.

Experimental investigation of quantum discord has now been extended to solid systems. In a recent experiment, a system of electron and nuclear spin in a phosphorous donor in silicon was observed by generating a series of separable thermal states with different anisotropic parameters\cite{Rong2012}. The authors observed nonzero quantum discord and found that there were sudden change behaviors of the quantum discord during tuning of the anisotropic parameters. There are also other experimental works concerning total quantum correlations; e.g. a modified correlation witness\cite{Aguilar2012,Passante2011} and the space of quantum correlations\cite{Chiuri2011}. These types of systems could provide an intriguing perspective on the dynamics of quantum discord, and help to better determine different experimental technologies.

\section{Conclusion}

Past investigations of quantum discord in open quantum systems have helped to disclose the particular dynamical behaviors of discord, as well as applications of both fundamental and practical importance. Nevertheless, observations of Markovian zero-discord classicality\cite{Arsenijevic2012}, and the unusual response of quantum discord to temperature\cite{Freitas2012}, still show that further investigations on the quantifications of quantum and classical correlations are required. Another arresting aspect is that quantum discord can be created under local operations\cite{Streltsov2011_2,Hu2012,Ciccarello2012}, which is different from that of entanglement\cite{Horodecki2009}. Gessner {\it et al.} suggested\cite{Gessner2012} that quantum discord should be interpreted as the local measure of quantumness, not necessarily as the number of quantum correlations. As a result, the problem of classifying quantum and classical correlations is still open despite the existence of the many different ways to measure nonclassical correlation introduced in Part 2.

From the above theoretical and experimental investigations, we can see that the time evolutions of both quantum discord and classical correlation in open quantum systems can be deduced from the time evolution of the initial states by using quantum state tomography and optimization calculations (or the corresponding witness circuits). Inspired by the factorization law of entanglement evolution in noisy quantum channels\cite{Konrad2008,Farias2009,Xu2009}, we expect that a similar simple relationship exists to characterize quantum correlation in a noisy environment. It has already been shown that any quantum correlation should be related to the amount of entanglement\cite{Streltsov2011}. Therefore, suitable quantification of nonclassical correlations is needed to meet this mission.

More experimental works must be done to further understand quantum discord, especially to clarify the quantum advantages in quantum information processing and to demonstrate their fundamental applications. However, increasing interest in this type of work will certainly lead to more distinctive discoveries in this field.

\section*{Acknowledgements}

This work is supported by the National Basic Research Program of China (Grants No. 2011CB9212000), National Natural Science Foundation of China (Grant Nos. 11004185, 60921091, 10874162), the Fundamental Research Funds for the Central Universities (Grant No. WK 2030020019).

\end{document}